\documentclass[aps,prd,reprint,superscriptaddress,preprintnumbers,showpacs,nofootinbib]{revtex4-1}

\usepackage[colorlinks, pdfborder={0 0 0}, plainpages=false]{hyperref}
\usepackage[utf8]{inputenc}
\usepackage{graphicx}
\usepackage{amsmath}
\usepackage{amssymb}
\usepackage{acronym}
\usepackage{amsfonts,pifont}
\usepackage{txfonts}
\usepackage[normalem]{ulem}
\usepackage{booktabs}
\usepackage{multirow}
\usepackage{soul}

\usepackage{etoolbox}
\newrobustcmd{\icon}[1]{\includegraphics[height=12pt]{#1}}

\newcommand{\cardiff}{School of Physics and Astronomy, Cardiff University, The
Parade, Cardiff, CF24 3AA, United Kingdom}

\newacro{BH}{Black Hole}
\newacro{BBH}{Binary Black Hole}
\newacro{NR}{Numerical Relativity}
\newacro{GW}{Gravitational Wave}
\newacro{IMR}{Inspiral-Merger-Ringdown}
\newacro{NS}{Neutron Star}
\newacro{BNS}{Binary Neutron Star}
\newacro{LSC}{LIGO Scientific Collaboration}
\newacro{LVC}{LIGO-Virgo Collaboration}
\newacro{LVK}{LIGO-Virgo-KAGRA Collaboration}
\newacro{LIGO}{Laser Interferometric Gravitational-Wave Observatory}
\newacro{aLIGO}{Advanced LIGO}
\newacro{Adv}{Advanced Virgo}
\newacro{EM}{electromagnetic}
\newacro{PN}{Post Newtonian}
\newacro{EOB}{Effective-One-Body}
\newacro{SPA}{stationary phase approximation}
\newacro{SNR}{signal-to-noise ratio}
\newacro{PSD}{power spectral density}
\newacro{ROM}{reduced order model}
\newacro{ROQ}{reduced order quadrature}

\newacro{ANN}{artificial neural network}

\begin{document}

\preprint{LIGO-P2300455}

\title{Probabilistic Model for the Gravitational Wave Signal from Merging Black Holes}

\author{Sebastian Khan}
\affiliation\cardiff


\date{\today}

\begin{abstract}
Parameterised models that predict the gravitational-wave (GW) signal from
merging black holes are used to extract source properties from GW observations.
The majority of research in this area has focused on developing methods capable
of producing highly accurate, point-estimate, predictions for the GW signal.  A
key element missing from every model used in the analysis of GW data is an
estimate for how confident the model is in its prediction.  This omission
increases the risk of biased parameter estimation of source properties.
Current strategies include running analyses with multiple models to measure
systematic bias however, this fails to accurately reflect the true uncertainty
in the models. In this work we develop a probabilistic extension to the
phenomenological modelling workflow for non-spinning black holes and
demonstrate that the model not only produces accurate point-estimates for the
GW signal but can be used to provide well-calibrated local estimates for its
uncertainty.  Our analysis highlights that there is a lack of Numerical
Relativity (NR) simulations available at multiple resolutions which can be used
to estimate their numerical error and implore the NR community to continue to
improve their estimates for the error in NR solutions published.  Waveform
models that are not only accurate in their point-estimate predictions but also
in their error estimates are a potential way to mitigate bias in GW parameter
estimation of compact binaries due to unconfident waveform model
extrapolations.
\end{abstract}

\maketitle

\section{Introduction}
\label{sec:intro}

The strongest
gravitational-wave (GW) signals contain the most information about the source
that produced them. In order to maximise the amount of science we can extract
from GW signals we must build detailed physical models that describe how
compact binaries merge. Waveform models are the culmination of the efforts of
the community who research new modelling techniques~\cite{Thompson:2023ase,
Pompili:2023tna, Ramos-Buades:2023ehm, PhysRevD.104.124027, Yu:2023lml,
Khalil:2023kep, Islam:2022laz, Jaramillo:2022mkh, PhysRevD.105.044035,
Nagar:2023zxh,Edwards:2023sak, McWilliams:2018ztb,2023arXiv231111311A,
2023arXiv231016980G} to accurately and efficiently include all the relevant
physical effects that are predicted to be important for the current generation
of ground based GW detectors.

However, whilst the loudest events have the most scientific potential they are
also the most susceptible to systematic and statistical errors in waveform
models that can bias information extraction or masquerade as deviations of
General Relativity~\cite{Hu:2022bji}. As detectors continue to be improved,
reaching higher levels of sensitivity, studies have shown that current
numerical relativity codes and waveform models are not yet accurate
enough~\cite{Samajdar:2018dcx, Gamba:2020wgg, Thompson:2020nei,
PhysRevResearch.2.023151, Rashti:2023wfe} to minimise the impact of systematic
errors. Indeed, waveform systematics have already begun to impact current
analyses~\cite{Hannam:2021pit, Khan:2019kot, Chatziioannou:2019dsz}.

Estimating and modelling waveform error is a growing area of research with
several methods proposed that can reduce the impact of waveform model
systematic error on GW parameter estimation.  Methods such
as~\cite{PhysRevD.101.064037, PhysRevD.106.083003,
Ashton:2021cub,Puecher:2023rxw} perform parameter estimation with multiple
models either simultaneously or separately and combine their posterior samples
according to their Bayesian evidence.  These types of methods currently only
account for the relative error between models and do not consider the
accuracy of each model.  The following methods take a waveform modelling
approach and require access to Numerical Relativity (NR) data in the region of
parameter space of interest.  The first method assumes you have an existing
model which you use as a baseline.  First you construct the residual between
the baseline model and NR which is subsequently modelled using Gaussian process
regression (GPR). This can be utilised in Bayesian parameter estimation by a
modified likelihood function that marginalises over the uncertainty of the
residual model~\cite{PhysRevLett.113.251101, PhysRevD.91.124062, Moore:2015sza}.  Another similar
method~\cite{PhysRevD.101.063011} proposes to build a GPR model by directly
interpolating NR data.  Recently, it has been suggested to introduce waveform
systematic uncertainty into waveform models as frequency-dependent amplitude
and phase corrections in a similar procedure to how detector calibration
uncertainty is included and subsequently marginalise over these corrections in
parameter estimation~\cite{Read:2023hkv}.

We approach this problem from a waveform modelling perspective and explicity
build a parametric phenomenological fit calibrated to NR solutions.  By using
multiple NR waveforms of different numerical resolutions and from different
numerical codes we estimate the NR uncertainty which feeds directly into our
model. Our fit to discrete individual NR waveforms is then extended into a
continuous model using non-parametric GPR that endows the model with a number
of desireable properties. The first is that it naturally provides a measure of
uncertainty. Second, with an appropriate choice of kernel, the uncertainty
grows as you move away from training points, this gives the model a sense for
when you are evaluating the model in regions where it has not been constrained.
Our model is a semi-parametric probabilistic model for the GW signal from
merging black holes that not only provides a best-fit point estimate but can
explicitly produce waveform samples.  Similarly to~\cite{PhysRevD.96.123011} we
propose to use the difference between the best-fit waveform and a number of
randomly drawn waveform samples produced from our model to estimate the true
error between the best-fit and the NR solution.

Our method extends existing phenomenological approaches which have already been
developed to accurately model a wide range of compact binary coalescences
(BBH~\cite{Husa:2015iqa, PhysRevD.93.044007, PhysRevLett.120.161102,
PhysRevD.100.024059, Khan:2019kot, PhysRevD.104.124027, PhysRevD.102.064001,
Estelles:2020osj, Estelles:2020twz, PhysRevD.103.104056},
BNS~\cite{PhysRevD.99.024029, PhysRevD.100.044003},
NSBH~\cite{Thompson:2020nei}) and will empower these models with the ability to
quantify their confidence in their predictions.  Probabilistic waveform models
that are not only accurate but have accurate error estimation is crucial for
Bayesian parameter estimation methods that marginalise over waveform
uncertainty and will safeguard GW astronomy against overly confident
extrapolations.

\begin{figure*}[tbh] 
\includegraphics[width=\textwidth]{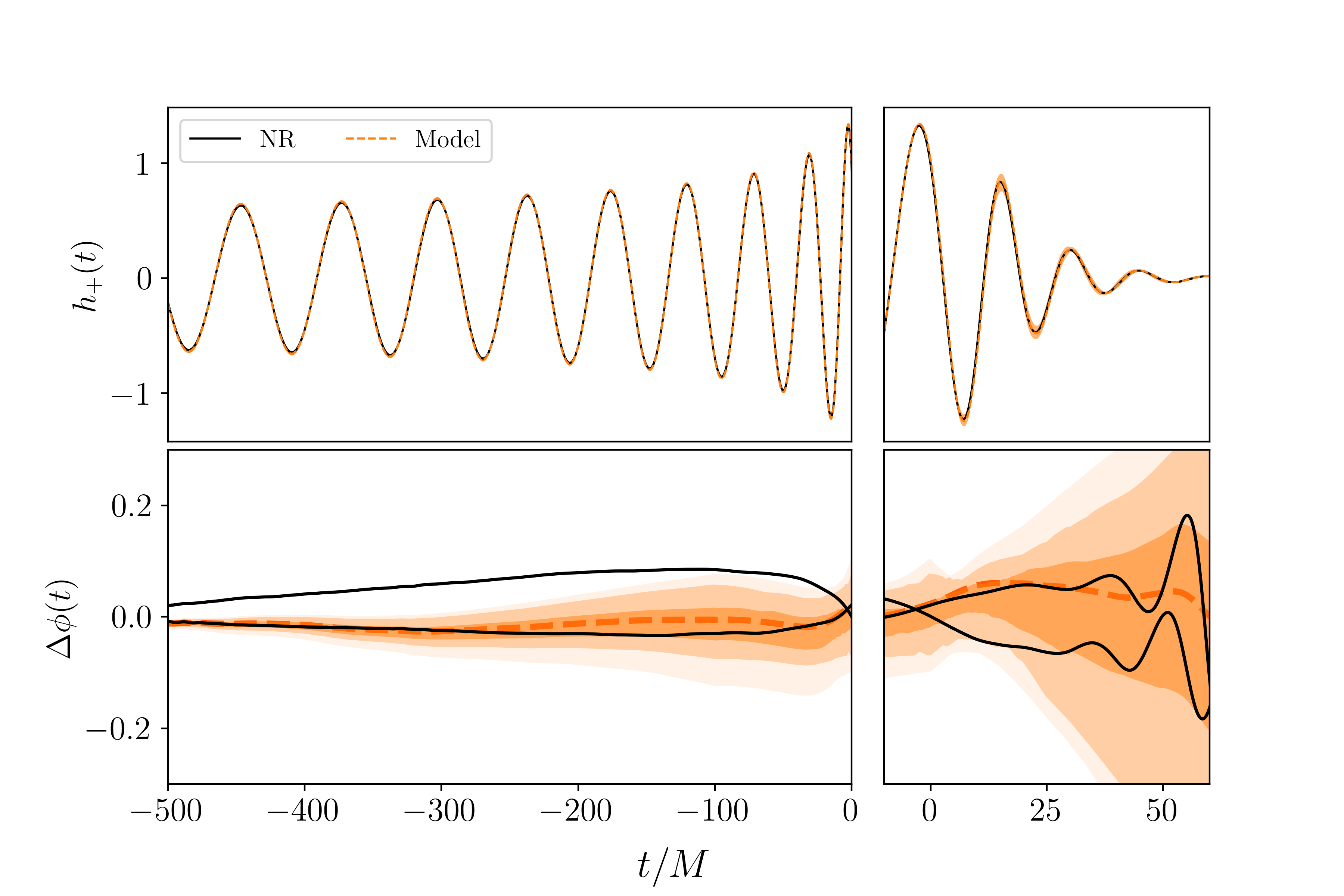}
\caption{
GW signal from a mass-ratio 8:1 binary black hole system compared with
predictions from the PPM model. This NR simulation was not used to train the
model. Top row: $h_+$ from the highest resolution NR simulation (black), the
mean PPM model prediction (dashed orange) and the $90^{th}$ percentile width
from 100 PPM samples (orange shaded region). Top left panel: shows the inspiral
up to $t=0M$. The level of accuracy and uncertainty are such that deviations
between the NR and the model are barely visible on this scale. Top right panel:
shows the ringdown where uncertainty in the model is more noticeable. Bottom
row: the phase error. The black lines are the two lower resolution NR
simulations compared with the reference NR simulation. The orange dashed line
is the PPM mean and the orange shaded regions show the $50^{th}$, $90^{th}$ and
$99^{th}$ percentile width of the phase error distribution from 100 PPM
samples. The match between the highest and lowest NR simulation is $0.9996$.
The match between highest resolution NR simulation and the PPM mean is
$0.99984_{0.99936}^{0.99993}$. Where the upper and lower bounds are the
$5^{th}$ and $95^{th}$ percentile of the match between 100 PPM samples and the
NR simulation.
}
\label{fig:wavephase}
\end{figure*}

In Figure~\ref{fig:wavephase} we show an example of how our new Probabilistic
Phenomenological Model (PPM) can generate waveform samples as well as
the mean waveform. We compare against 3 NR simulations of a
mass-ratio 8:1 binary black hole system. The match between the PPM mean and NR
ranges from $0.9994$ to $0.99994$ depending on which simulation we compare
with. In the top row we show the $h_+$ polarisation optimised over a relative
time and phase shift. In the lower panel we plot the phase difference. The
black lines are the phase difference between the NR waveforms. The orange
dashed line is the phase difference between the reference NR simulation and the
PPM mean prediction. The orange shaded regions show the $50^{th}$, $90^{th}$ and
$99^{th}$ percentile width of the phase error distribution from 100 PPM
samples. For this case the only visible variance in the PPM model can be seen
during the ringdown in the top right panel.

In the remainder of this paper we describe our methodology and demonstrate the
model's accuracy.

\section{Preliminaries}
\label{sec:prelim}

We consider a binary black hole system with masses $m_1$ and $m_2$.
Their mass-ratio is defined as $q = m_1 / m_2 \geqslant 1$ and the
symmetric mass-ratio is $\eta = m_1 m_2 / M^2$ where $M = m_1 + m_2$ is the total mass.
The complex GW strain is defined as

\begin{equation}
h(t; q, \theta, \phi) = h_+ - i h_\times = \sum\limits_{\ell, m} h_{\ell m}(t; q) \,  {}^{-2}Y_{\ell m}(\theta, \phi) \, .
\end{equation}

The angular dependency is factored out using spin-weight $-2$ spherical
harmonics reducing the waveform to a one-dimensional timeseries which is a
function of the physical parameters, only mass-ratio in this case.

By restricting the type of binary black hole system we are modelling to
non-spinning black holes with negligible orbital eccentricity we can
approximate the full GW complex strain with just the
$(\ell, m) = (2,\pm2)$. Futhermore, due to the fixed orbital plane
the positive $m$ and negative $m$ multipoles are related to each other
via a complex conjugate. Here we choose to model the $h_{22}$ multipole.
This approximation deteriorates as the mass-ratio increases
because the relative amplitude of
$(\ell, m) \neq (2,\pm2)$ also typically increases.

We can decompose the complex multipole into an amplitude and a phase
which is a successful method to describe and model binary merger evolution
with analytical models, this decomposition is defined as

\begin{equation}
h_{22}(t) = A_{22}(t) e^{-i \phi_{22}(t)} \, .
\end{equation}

We choose to directly model the angular GW frequency
$\omega_{22} (t) := d\phi_{22}(t)/dt$
and then integrate this to obtain the GW phase.

The ringdown of the remnant black hole is described
analytically as a superposition
of quasi-normal modes. The ringdown angular frequency
is $\omega_{RD} = 2\pi f_{RD}$ and the angular damping time
is $\tau_{damp} = 2\pi t_{damp}$.
$f_{RD}$ and $t_{damp}$ are expressed as functions of the
mass-ratio of the binary and we use model developed in~\cite{Husa:2015iqa} here.
The amplitude of the ringdown is not analytically known and is a quantity
that we explicitly model.
Here we have omitted indicies
$\ell$, $m$ and $n$ which indicate which multipole and overtone
ringdown mode is being considered however, we only model the $(\ell, m, n) = (2, 2, 0)$
multipole.

\section{Data}
\label{sec:data}

In this section we descibe the numerical relativity dataset we have aggregated
across several code bases.  We use NR solutions from four different NR groups
namely; SXS catalogue~\cite{SXSCATALOGUE}, GTech/UTexas
catalogue~\cite{GTCATALOGUE}, RIT catalogue~\cite{RITCATALOGUE} and BAM.  The
BAM simulations used here are not publicly available currently however, there
is a public catalogue of precessing simulations~\cite{Hamilton:2023qkv}.  To
convert the BAM $\psi_4$ data to strain we used the software package
POWER~\cite{Johnson:2017oel}.

Our dataset consists of 25 unique mass-ratios ranging from $q=1$ up to $q=32$.
10 of the simulations have more than one NR simulation either performed by
different codes or the same code but at a different numerical resolution.  In
table~\ref{table:nrsims} we list the NR simulations we use.  Assessing the
accuracy of an NR simulation can be challenging.  You typically need to perform
at least 3 NR simulations at varying levels of numerical resolution in
order to perform a convergence test. Even then the results of a convergence
test can be difficult to interpret due to the sophistocated and complex
numerical methods used.  See~\cite{Hamilton:2023qkv} for a recent NR catalogue
analysis.

Comparisons within the same code base can test the accuracy of the code
however, there could exist systematic code errors~\cite{Boyle:2015nqa} that are
easier to detect by comparing with an independent NR code.  The difficulty with
cross-code comparisons is that it is not necessarily possible to prefer one
solution over another (without the results of a convergence test).  This is the
case for the majority of NR solutions available.
Additionally, some NR simulations have been superseded by more accurate ones and
therefore these simulations are not necessarily representatative of the accuracy of
current NR codes.
In this study we have intentionally used NR solutions from not only different code
bases but also using multiple simulations performed at different numerical
resolutions in order to test our method to build a model that can estimate the
NR error. Throughout we will assume all the NR simulations are equally accurate
which is a potential cause of bias in our results.

The length of each NR simulation is highly varied.  The majority of NR
simulations are between $\sim 900 M$ and $\sim 2000 M$ long. These simulations
are not long enough on their own to build and test an inspiral model however,
they are long enough to develop the modelling workflow. Typically, short NR
simulations are hybridised with \ac{PN} inspiral waveforms to achieve the
desired length.  In order to include as many NR simulations as possible we
truncate all NR simulations to a length of $800 M$ which, takes into account
removing of an initial $140 M$ of junk radiation and keeping $\sim 90 M$ of the
ringdown signal.

As we will describe in the next section, we use a collocation fitting algorithm
where the coefficients of the model are values of the data at various points in
time.  We first align our data such that the peak of the amplitude is at $t =
0M$.  To facilitate comparison between NR simulations with the same parameters
we apply an additional time and phase shift that minimises the phase error
between NR simulations over the first $800M$.

\addtolength{\tabcolsep}{2pt}
\begin{table*}
\begin{tabular}{rrllrrllllll}
\hline
\hline
\toprule
\# &    q &               name &  code & \# &    q &               name &  code & \# &     q &                 name &  code \\
\hline
\midrule
    1 & 1.00 & RIT-eBBH-1090-n100 & LazEv &    20 & 2.25 &             GT0757 &  Maya &  39 &   7.0 &    RIT-BBH-0416-n140 & LazEv \\
    2 & 1.00 &  RIT-BBH-0112-n100 & LazEv &    21 & 2.35 &             GT0380 &  Maya &  40 &   8.0 &  q8a0a0\_T\_96\_504n512 &   BAM \\
    3 & 1.00 &  SXS\_BBH\_0180\_Res4 &  SpEC &    22 & 2.41 &  RIT-BBH-0139-n140 & LazEv &  41 &   8.0 &   q8a0a0c05\_T\_80\_420 &   BAM \\
    4 & 1.00 &  SXS\_BBH\_0180\_Res2 &  SpEC &    23 & 2.50 &             GT0565 &  Maya &  42 &   8.0 & q8a0a0\_T\_112\_588n768 &   BAM \\
    5 & 1.00 &  SXS\_BBH\_0180\_Res3 &  SpEC &    24 & 3.00 &             GT0453 &  Maya &  43 &  10.0 &    SXS\_BBH\_0303\_Res4 &  SpEC \\
    6 & 1.18 &  RIT-BBH-0084-n100 & LazEv &    25 & 4.00 &             GT0454 &  Maya &  44 &  10.0 &    RIT-BBH-0978-n144 & LazEv \\
    7 & 1.20 &             GT0898 &  Maya &    26 & 4.00 &  SXS\_BBH\_0167\_Res5 &  SpEC &  45 &  10.0 &    SXS\_BBH\_0303\_Res5 &  SpEC \\
    8 & 1.25 &             GT0738 &  Maya &    27 & 4.00 &      q4a0\_T\_80\_320 &   BAM &  46 &  10.0 &    SXS\_BBH\_0303\_Res3 &  SpEC \\
    9 & 1.33 & RIT-eBBH-1241-n100 & LazEv &    28 & 4.00 & RIT-eBBH-1133-n100 & LazEv &  47 &  10.0 &    q10c25e\_T\_112\_448 &   BAM \\
   10 & 1.50 &             GT0477 &  Maya &    29 & 4.00 &  SXS\_BBH\_0167\_Res3 &  SpEC &  48 &  15.0 &    RIT-BBH-0957-n084 & LazEv \\
   11 & 1.75 &             GT0727 &  Maya &    30 & 4.00 &      q4a0\_T\_96\_384 &   BAM &  49 &  15.0 &    RIT-BBH-0373-n140 & LazEv \\
   12 & 1.82 &  RIT-BBH-1020-n144 & LazEv &    31 & 4.00 &     q4a0\_T\_112\_448 &   BAM &  50 &  15.0 &    RIT-BBH-0942-n120 & LazEv \\
   13 & 2.00 &  SXS\_BBH\_0169\_Res3 &  SpEC &    32 & 5.00 &  SXS\_BBH\_0107\_Res3 &  SpEC &  51 &  18.0 &  q18a0a0c025\_96\_fine &   BAM \\
   14 & 2.00 &  SXS\_BBH\_0169\_Res4 &  SpEC &    33 & 5.00 &  RIT-BBH-0152-n120 & LazEv &  52 &  18.0 &      q18a0a0c025\_120 &   BAM \\
   15 & 2.00 &  SXS\_BBH\_0169\_Res5 &  SpEC &    34 & 5.00 &             GT0577 &  Maya &  53 &  18.0 &      q18a0a0c025\_144 &   BAM \\
   16 & 2.00 & RIT-eBBH-1200-n100 & LazEv &    35 & 5.00 &  SXS\_BBH\_0107\_Res4 &  SpEC &  54 &  32.0 &    RIT-BBH-1025-n100 & LazEv \\
   17 & 2.00 &             GT0446 &  Maya &    36 & 5.00 &  SXS\_BBH\_0107\_Res5 &  SpEC &  55 &  32.0 &    RIT-BBH-0792-n120 & LazEv \\
   18 & 2.05 &             GT0378 &  Maya &    37 & 6.00 &             GT0604 &  Maya &       &       &                      &       \\
   19 & 2.20 &             GT0379 &  Maya &    38 & 6.00 &  RIT-BBH-0090-n100 & LazEv &       &       &                      &       \\
\hline
\hline
\bottomrule
\end{tabular}
\caption{Numerical Relativity Simulations Used.
We used $q \in \{1, 2, 5, 6, 10, 18\}$ for training and the rest
for testing.}
\label{table:nrsims}
\end{table*}
\addtolength{\tabcolsep}{-2pt}

\section{Method}
\label{sec:method}

The modelling process is split into two main steps: i) a parametric part and ii)
a non-parametric part.  Schematically, the parametric ansatz is a function of
time $t$ and is parameterised with parameters $\theta$ i.e. $f(t; \theta)$ with
$\theta$ being determined by fitting the ansatz to the data.  The $\theta$
coefficients are then expressed as a function of the mass-ratio $q$ which we
construct using a non-parametric function $g(q)$.  We write our semi-parametric
model as an approximation of some target function $y$ as

\begin{equation}
    y(t; q) \approx f(t; g(q)) \, .
\end{equation}

Here our target functions are the amplitude and angular frequency of the $(\ell,
m) = (2, 2)$ multipole.  Some of the functional forms we use for our parametric
model are inspired by the work of~\citet{Estelles:2020osj, Estelles:2020twz}.

One of the motivating factors to pursuing this approach was to build an
interpretable model. The more interpretable a model is the easier it is for a
practitioner to understand how the model produced the output it did. A high
degree of interpretability is easiest to obtain for linear models. As such we
have attempted to build a model based purely on linear ans{\"a}tze. With a
linear model we also have the ability to use the collocation fitting algorithm
in which the coefficients of the model are values of the data at specific
points in time (the collocation points). The coefficients $\theta$ of the
ansatz are obtained by solving a linear system of equations at the time of
inference. By modelling directly the value of the waveform we typically find
smoother samples to interpolate (when fitting the non-parametric part of the
model) and we also gain interpretability because now the error in the
coefficients corresponds to the error in either the amplitude or the frequency
at the specific points in time. If we assume the model coefficients are
independent then the uncertainty in model coefficients can be directly read off
of the data as opposed to estimating the covariance matrix. This method can
work for non-linear functions by first finding optimal values for the
non-linear coefficients, essentially treating them as hyper-parameters. After
the optimal values have been found they can be fixed which transforms the
non-linear ansatz into a linear ansatz. For some ans{\"a}tze
the model coefficients could have significant correlations between them,
in this case it might be necessary to map out the posterior distribution
using markov chain monte carlo sampling techniques. In these cases, it will
be more complicated to construct accurate waveform samples as it will require
a model for the joint distribution.

Once the collocation values have been extracted from the discrete dataset we
build a continuous model for them as a function of the physical parameters,
just the mass-ratio in this case. There are many methods to do this for example
using polynomials or artificial neural networks~\cite{PhysRevD.103.064015}.
Here we use Gaussian process regression (GPR) which has been used in models for
aligned-spin BBH surrogates~\cite{PhysRevD.96.123011, Varma:2018mmi}, to model
the BH remnant properties~\cite{Varma:2018aht} and even a prototype 7D
precessing model~\cite{PhysRevD.101.063011}.  Gaussian Processes (GPs) have
recently been used to model transient noise events (also called glitches) in
GW detector data~\cite{Ashton:2022ztk} as well as for density
estimation~\cite{DEmilio:2021laf}.

We have explored a blend of parametric and non-parametric methods to build a
semi-parametric model that combines desireable qualities from both methods.  We
use a parametric model to describe waveform phenomenology.  This gives the
model a strong underlying physical structure for example, the frequency of
non-spinning BBHs is monotonic. A physical constraint such as this is not
necessarily imposed in a non-parametric model (however, it is possible to
impose such constraints).  In fact due to the specifics of our model, if the
errors in the coefficients are large then this monotonicity can be broken
however, this should be in regions where the model uncertainty is also large.
We then switch to using a non-parametric model to fit the coefficients of the
parametric models as a function of the physical parameters.  A non-parametric
approach is optimal here because we have less physical intuition about the
phenomenology of how these coefficients should behave and we can leverage the
power of a method like GPR which is a flexible model (i.e., can typically fit
the data well) and naturally provides a local measure of uncertainty.

\section{Parametric Model}
\label{sec:para-model}

\subsection{Collocation Method}
\label{sec:collocation}

For our parametric model we use the collocation method to solve our linear
regression problem. In standard least-squares regression the practitioner
proposes an ansatz with $N$ coefficients ($\boldsymbol{\theta}$) which are
determined by minimising the least-squares error between the model and
data.  In the collocation method we solve the same problem of fitting an ansatz
to the data however, we have more control over properties of the solution, for
example, we can additionally constrain the value of the derivative of the
ansatz at particular times.  For an ansatz with $N$ coefficients we first
specify a set of $N$ \emph{collocation points}, $\{P\}$.  Second, we evaluate
the data and/or the $n^{th}$ derivative of the data at the collocation points
which we call the set of \emph{collocation values} $\{V\}$.
The coefficients of the ansatz are computed by solving the
following linear system of equations

\begin{equation}
\boldsymbol{\mathcal{I}} \boldsymbol{\theta} = \mathbf{V} \, ,
\label{equ:collocation}
\end{equation}

Where we define $\boldsymbol{\mathcal{I}}$ as the \emph{information matrix}.
It's elements are the values of the variables (also called indeterminates) of
the ansatz evaluated at each of the collocation points. We implemented our
collocation method using the SymPy python library~\cite{10.7717/peerj-cs.103}
to perform symbolic differentiation.

We illustrate this method with a simple 1D regression example.
Suppose our discrete data are $\{X_i, Y_i\}$ and we have approximated this data with
an interpolating function $y(x)$. For our ansatz we use a quartic polynomial.

\begin{equation}
f(x; \theta) = \sum\limits_{i=0}^{i \le 4} \theta_i x^{i} \, .
\end{equation}

We select as our collocation points $p = \{ 0, 0.5, 1 \}$, $p' = \{ 0, 1 \}$
and define $P = p \cup p'$, we use a prime on the variable to represent the
derivative order at which the collocation points should be evaluated at.  These
collocation points represent constraining the value of the ansatz and it's
first derivative at the boundaries and then constraining the ansatz at the
midpoint.  The collocation values are therefore $v = \{ y(0), y(0.5), y(1) \}$
and $v' = \{ \frac{dy}{dx}(0), \frac{dy}{dx}(1) \}$. From this we collect together
the collocation values into a vector $V = v \cup v'$.
If we explicitly write out the matrix equation for (Equation~\ref{equ:collocation})
for this system we get

\begin{equation}
\begin{bmatrix}
1 & 0 & 0 & 0 & 0\\
1 & 1/2 & 1/4 & 1/8 & 1/16\\
1 & 1 & 1 & 1 & 1\\
0 & 1 & 0 & 0 & 0\\
0 & 1 & 2 & 3 & 4
\end{bmatrix}
\begin{bmatrix}
    \theta_{0} \\
    \theta_{1} \\
    \theta_{2} \\
    \theta_{3} \\
    \theta_{4}
  \end{bmatrix}
=
\begin{bmatrix}
    V_{0} \\
    V_{1} \\
    V_{2} \\
    V_{3} \\
    V_{4}
  \end{bmatrix} \, .
\end{equation}

\begin{figure*}[tbh]
\includegraphics[width=\textwidth]{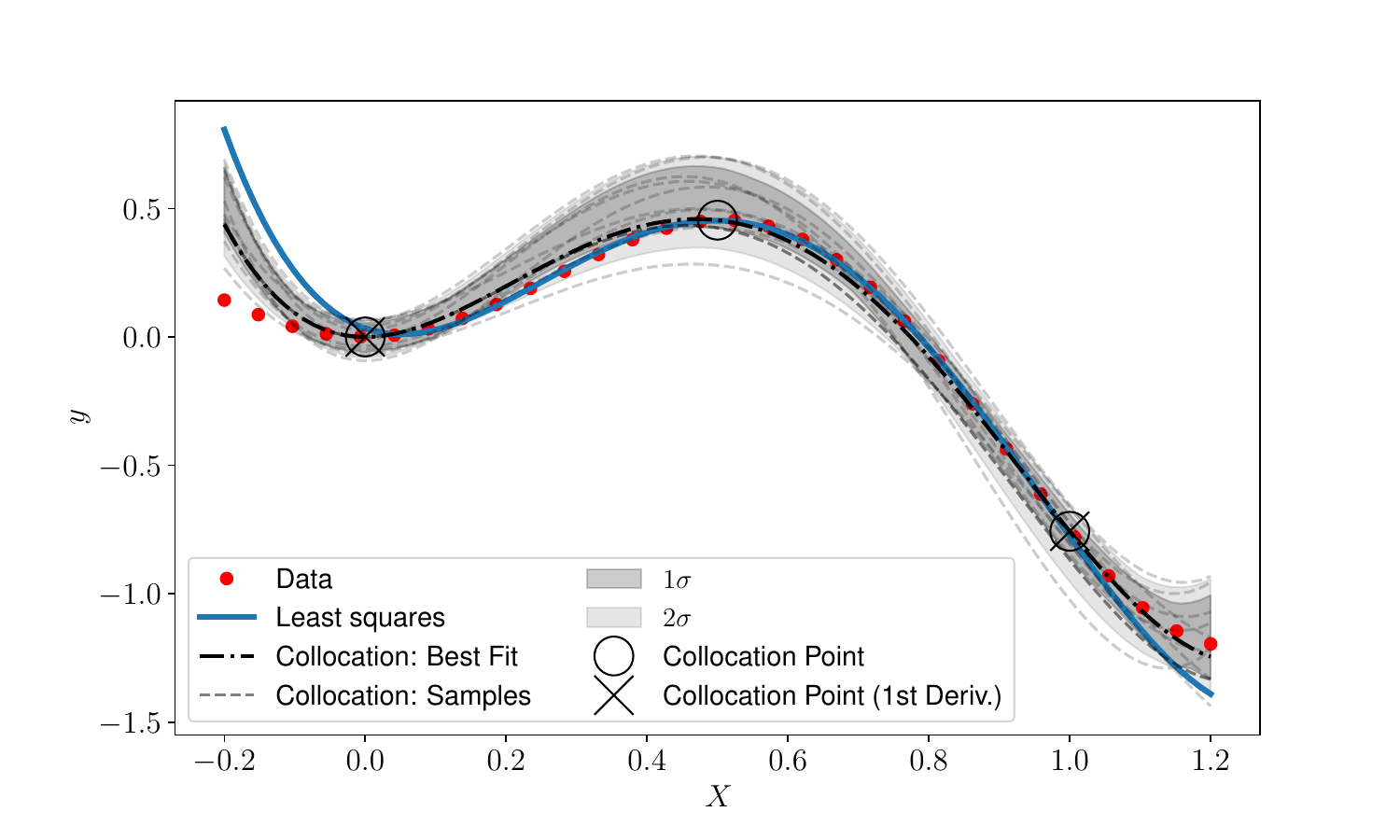}
\caption{Toy example to illustrate collocation method. We compare the least-squares approach (blue)
and the collocation point approach (black/grey) when applied to the task of modelling the data (red).
For the collocation point method with specify the value and the first derivative at $x=0$ and $x=1$ as well as
a collocation point at $x=0.5$. To generate samples from the collocation point method we perturb the
fitted collocation point values with a $\mathcal{N}(0,0.1)$ distribution}
\label{fig:collocation_example}
\end{figure*}

We solve this system of equations for $\boldsymbol{\theta}$ at the time of inference.
In Figure~\ref{fig:collocation_example} we compare the least squares approach with the
collocation method for a simple toy function $y(x) = x^2 \sin(4x)$. Note that we show $x \notin [0,1]$
to illustrate how this particular model extrapolates outside the training set.
Over the training set the least squares fit has the smallest error however, it does not
necessarily match the boundary well which is most easily seen at $x=0$. On the other hand,
the collocation method, with $0^{th}$ and $1^{st}$ derivative constraints at the boundaries
is guaranteed to fit the data within the numerical accuracy used. We also show how we can
easily perturb the $V$ vector around their true value in an interpretable way to produce
\emph{samples}. Specifically, to each element of $V$ we add a random sample from a $\mathcal{N}(\mu=0, \sigma=0.1)$
distribution to simulate uncertainty in our fit of the $V$ vector.
The equivalent method for least squares is to add perturbations according to
the covariance matrix for fit.

\subsection{Frequency model}
\label{sec:freq_model}

In what follows we use a caret ($\hat{\cdot}$) to indicate a fitted quantity.
We split the frequency into three regions which we call the inspiral
$\omega_{I}(t) \in [-700, -100]M$, merger $\omega_{M}(t) \in [-100, 0]M$,
ringdown $\omega_{R}(t) \in [0, 87]M$. The times define the regions used for
fitting and testing the model.

The inspiral model is written as a correction to the TaylorT3
approximant. We generate the $(\ell, m) = (2,2)$ GW angular frequency,
denoted as $\omega_{22}(t)$,
using a 3.5 PN order accurate expression for non-spinning binaries
~\cite{Blanchet:2001ax, Buonanno:2009zt} written as

\begin{eqnarray}
    \omega_{22}(t) &=& \omega_N(t) \, \sum\limits_{k=0}^{7} \omega_k \Theta^{k} \, , \\
    \omega_{orb}(t) &=& \omega_{22}(t) / 2 \, .
\end{eqnarray}

Where $\Theta(t) = \left(\frac{\eta}{5M}(t_c-t)\right)^{-1/8}$, $\omega_k$ are
expansion coefficients~\cite{Buonanno:2009zt} and the leading order Newtonian
term is given $\omega_N(t) = \Theta^3(t) / 8$.  $t_c$ is the time which the
TaylorT3 expansion formally diverges in what follows we set this value to $t_c =
0$.  Additionally, $\omega_{orb}(t)$ is the orbital angular frequency which
is an input for the inspiral amplitude model.

We model the residual between the PN and NR GW angular frequency, factoring out $\omega_N$

\begin{equation}
\omega_{res}(t) = \frac{\omega_{NR}(t) - \omega_{22}(t)}{\omega_N(t)} \, .
\end{equation}

The residual $\omega_{res}(t)$ is evaluated at the following
collocation points $T_{\omega}^{I} = \{ -700, -300, -100\} M$.
Our ansatz to fit this model is given by the next 3 terms in the PN series
and therefore extending the PN model to pseudo-5 PN order given by

\begin{equation}
    \hat \omega_{res}(t) = \sum\limits_{i=0}^{i \leqslant 2} \theta^{\omega_{res}}_i \Theta^{8+i} \, .
\end{equation}

The model prediction for the inspiral is therefore,

\begin{equation}
    \hat \omega_{I}(t) = \omega_{22}(t) + \omega_N(t) \, \hat\omega_{res}(t) \, .
\end{equation}

We call the region between the end of the inspiral and the beginning
of the ringdown the merger.
We found the ansatz proposed in~\cite{Estelles:2020osj} to be accurate and adopt
it here, as well as for the amplitude model in the next section.
The collocation points for this region are $T_{\omega}^{M} = \{ -100, -12, 0\} M$.
The merger ansatz is a power series in $\text{arcsinh}$ with a fixed width of $1/\tau$
(where $\tau$ is the damping time of the remnant black hole) given by

\begin{equation}
    \hat \omega_{M} (t) = \sum\limits_{i=0}^{i \leqslant 2} \theta^{\omega_M}_i \text{arcsinh}^{i}(t/\tau) \, .
\end{equation}

To model the ringdown portion of the waveform we found that a power series in $\tanh$
with a fixed wideth of $1/\tau$ works well.
The $\tanh$ function has a logistic shape which matches the phenomenology of the ringdown frequency well.
An improvement would be to explicitly include the ringdown frequency prediction from
perturbation theory, in a similar way to how the damping time $\tau$ is included.
We use the following collocation points $T_{\omega}^{RD} =\{-10, 0, 10, 40 \} M$
and the ansatz is given by

\begin{equation}
    \hat \omega_{RD} (t) = \sum\limits_{i=0}^{i \leqslant 3} \theta^{\omega_{RD}}_i \tanh^{i}(t/\tau) \, .
\end{equation}

In summary, the following set of frequency coefficients need to be fit
as a function of mass-ratio $\theta^{\omega} = \{\theta^{\omega_{res}} \cup \theta^{\omega_M} \cup \theta^{\omega_{RD}}\}$
where there is some redundency because the same collocation point is at the boundary between regions.
The final inspiral-merger-ringdown angular frequency function is defined piecewise as

\begin{equation}
\hat \omega_{IMR}(t) =
    \begin{cases}
    \hat \omega_{I}(t) & -700 M \leqslant t < -100 M \\
    \hat \omega_{M}(t)   & -100 M \leqslant t < 0 M \\
    \hat \omega_{RD}(t)  & 0 M    \leqslant t
    \end{cases}
\label{equ:omega_imr}
\end{equation}

\subsection{Amplitude model}
\label{sec:amp_model}

We split the amplitude into four regions which we call the
inspiral $A_{I}(t) \in [-700, -300]M$,
merger $A_{M}(t) \in [-300, 0]M$,
early-ringdown (ERD) $A_{ERD}(t) \in [0, 30]M$ and
late-ringdown (LRD) $A_{LRD}(t) \in [30, 87]M$.
For the merger and early-ringdown regions we scale the amplitude by $1/\eta$
which approximately removes the variability in the peak amplitude.
We decided to split the ringdown into and early- and a late- region
to allow us to use linear models and will be discussed below.

The inspiral model is written as a correction to the TaylorT3
approximant.
The amplitude of the $(\ell, m) = (2,2)$ mode is expressed as a function of
the PN parameter $x$ which is related to the orbital angular frequency
by the following relationship

\begin{equation}
x(t) = \omega_{orb}^{2/3}(t) \, .
\end{equation}

It is important to use the inspiral frequency model described in the previous
section to estimate $\omega_{orb}$ because the PN approximation can become
negative at late times causing $x$ to become complex. $x$ is therefore given by
$x(t) = (\hat\omega_I(t)/2)^{2/3}$.  The TaylorT3 PN inspiral amplitude is given
by

\begin{eqnarray}
A_{PN}(t) &=& A_N(t) \, \hat{H}^{22}(t) \, , \\
A_N(t) &=& 2\eta \sqrt{\frac{16 \pi}{5}} x(t)
\end{eqnarray}

Where $\hat{H}^{22}(t)$ is an expansion up to 3.5PN~\cite{Blanchet:2008je,
Faye:2012xt} and where we have defined an analogous Newtonian amplitude
pre-factor $A_N(t)$ which we will use to scale inspiral amplitude residuals by.
We first generate the TaylorT3 amplitude and construct the residual between that
and the NR data, scaled by $A_N$

\begin{equation}
A_{res}(t) = \frac{A_{NR}(t) - A_{PN}(t)}{A_N(t)} \, .
\end{equation}

Similarly to the inspiral frequency model we define our amplitude
inspiral ansatz as an extension of the PN model up to 4.5 PN order
given by

\begin{equation}
    \hat A_{res}(t) = \sum\limits_{i=0}^{i \leqslant 1} \theta^{A_{res}}_i x^{(8+i)/2}(t) \, .
\end{equation}

We only use two collocation points $T_{A}^{I} = \{ -700, -100\} M$
as we find that the majority of the amplitude data is explained well by
our inspiral frequency model $\hat{\omega_{I}}$.
The inspiral amplitude model is given by

\begin{equation}
    \hat A_{I}(t) = A_{PN}(t) + A_N(t) \, \hat A_{res}(t) \, .
\end{equation}

For the amplitude merger region, contrary to~\cite{Estelles:2020osj} who chose
an ansatz based on the $\text{sech}$ function, we use a power series in
$\text{arcsinh}$ with a width of $1/\tau$

\begin{equation}
    \hat A_{M} (t) = \sum\limits_{i=0}^{i \leqslant 3} \theta^{A_M}_i \text{arcsinh}^{i}(t/\tau) \, ,
\end{equation}

with collocation points given by $T_{A}^{M} = \{ -100, -50, -10, 0 \} M$.

The behaviour of the waveform after the peak of the amplitude is typically
called the ringdown region however, it is an active area of research to
determine the correct physics to describe the transtion the merger to the
ringdown~\cite{Giesler:2019uxc,Cook:2020otn,JimenezForteza:2020cve,Forteza:2021wfq}.
In recent years time domain waveform models tend to use the non-linear model
presented in~\cite{PhysRevD.90.024054} to model the ringdown region from the
peak amplitude onwards.  However, because the standard collocation point method
we use requires linear ans{\"a}tze we are unable to use this ringdown
parameterisation. Instead, we have introduced an ``early-ringdown'' region to
bridge the gap between the peak amplitude and the start of the ringdown region
which we loosely to be the times that can be accurately approximated by black
hole perturbation theory.  Motivated by the similarity between the onset and
falloff of the waveform around the peak we model the early-ringdown with the
same ansatz that we use for the merger amplitude. The early-ringdown ansatz is

\begin{equation}
    \hat A_{ERD} (t) = \sum\limits_{i=0}^{i \leqslant 4} \theta^{A_{ERD}}_i \text{arcsinh}^{i}(t/\tau) \, .
\end{equation}

The collocation points are $t_{ERD} = \{ 0, 5, 20, 30 \} M$ with an additional
collocation point evaluating the derivative of the peak amplitude $t_{ERD}' =
\{ 0 \} M$.  The full set of collocation points is therefore, $T_{A}^{ERD} =
t_{ERD} \cup t_{ERD}'$.  As we expect the derivative at the peak to be zero we
enforce this manually instead of fitting the collocation values for the
collocation point $t_{ERD}'$.

The late-ringdown ansatz is simply exponential decay with a decay constant
equal to the damping frequency of the remnant black hole $1/\tau$.

\begin{equation}
    \hat A_{LRD} (t) = \beta_{LRD} e^{-t/\tau}  \, .
\end{equation}

The constant $\beta_{LRD}$ is fixed by enforcing C(0) continuity and is defined as

\begin{equation}
    \beta_{LRD} = \hat A_{ERD} (t_0) e^{ t_0/\tau} \, .
\end{equation}

The matching time is a constant value of $t_0 = 30 M$. Defining a time after the
peak amplitude where the system can be fully described by perturbation theory is
an active area of research. For our purposes we need an approximate time after
which we can accurate transition to a purely exponential decay model.

In summary the following set of amplitude coefficients need to be fit
as a function of mass-ratio $\theta^{A} = \{\theta^{A_{res}} \cup \theta^{A_M} \cup \theta^{A_{ERD}}\}$.
The final inspiral-merger-ringdown amplitude function is defined piecewise as

\begin{equation}
\hat A_{IMR}(t) =
    \begin{cases}
    \hat A_{I}(t) & -700 M \leqslant t < -100 M \\
    \hat A_{M}(t)   & -100 M \leqslant t < 0 M \\
    \hat A_{ERD}(t) & 0 M    \leqslant t < 30 M \\
    \hat A_{LRD}(t) & 30 M   \leqslant t
    \end{cases}
\label{equ:amp_imr}
\end{equation}

\section{Non-parametric model}
\label{sec:nonpara-model}

In this section we describe our non-parametric model for
the parameter space fits needed to go from a discrete set of
data samples to a continuous model over the parameter space.

The target data is the set of all collocation values for both the amplitude and
frequency models described in the previous section. We gather the collocation
values together as $\theta = \{ \theta^{A} \cup \theta^{\omega} \}$.  For each
collocation value we will construct a non-parametric model as a function of the
mass-ratio i.e.  $g_{\theta}(q)$. To do this we will use the Gaussian process
regression algorithm.  The GPR algorithm begins by placing a Gaussian process
prior over the quantity of interest  written as

\begin{equation}
    g_{\theta}(q) \sim \mathcal{GP}(m_{\theta}(q), k_{\theta}(q,q')) \, ,
\end{equation}

with mean $m_{\theta}(q)$ and covariance function $k_{\theta}(q,q')$ for the
$\theta$ collocation point. Here, the GP model is simply multi-dimensional
Normal distribution with a covariance matrix constructed from the training set
according to a prescribed covariance function $k(q, q')$. We choose
$m_{\theta}(q) = 0$ and for the covariance function we use the Mat\'ern kernel
(with smoothness parameter $\nu = 5/2$)~\cite{books/lib/RasmussenW06}.

The kernel hyperparameters were determined by numerically
optimising the log marginal likelihood of the GP. We use the
scikit-learn~\cite{scikit-learn} implementation of Gaussian Process Regresssion
in our prototype model. A production-ready model would require either; a more
computationally performant GPR implementation, fast approximations such as
sparse-variational, random fourier features~\cite{NIPS2007_013a006f,
2020arXiv200209309W} or harware accelerators such as GPUs.

We found it necessary to transform the target variable to enforce the model
to make predictions that could not change sign, this could happen
when the target values are close to zero. For the amplitude and frequency
data this is a physical constraint. 
To constrain the model to only predict positve values we
exponentiate it's predictions and therefore we define the transformed
target variable as $z$ through the following equation

\begin{equation}
    z := \log(\,|\,y\,|\,) \, .
\end{equation}

Where $y$ is the target variable (i.e. collocation point values).
We can reverse this transformation so long as we keep track of the original sign of $y$.
Additionally, we also found that modelling $\log(q)$ helped to improve the extrapolation behaviour of the GP.

Next we discuss two types of uncertainty that our method accounts for.  For the
purposes of our fit of the collocation points we consider the variance between
collocation values at the same mass-ratio as the statistical (also called
aleatoric or data) uncertainty and the variance in regions devoid of training
data is called the systematic (also called epistemic or model) uncertainty of
the waveform model.

The statistical uncertainty is quantified by measuring the accuracy of NR
solutions (for example via a numerical convergence series) and can be reduced
by producing more accurate NR solutions.  The systematic uncertainty is a
measure of how well the model fit is constrained by the training data. To
reduce the systematic uncertainty you can perform new NR simulations at regions
where the model predicts large systematic uncertainty.  From the perspective of
NR it is known that simulations of high mass-ratio and/or rapidly rotating BHs
are typically much harder, numerically speaking, to simulate and therefore, it is
conceivable to expect the NR error to be larger in these regions of parameter
space.  In lieu of a full convergence series for each NR simulation in our
training set we take a conservative approach and assume each simulation is
equally accurate.  We use a homoskedastic noise model assuming a constant noise
variance which is an additional hyperparameter informed by the measured
variance in the training data.

The systematic uncertainty in GPR models can be controlled by the kernel
function.  Our choice of using a stationary kernel such as the Mat\'ern endows
the model with a notion of distance from the training set and as such can
produce models with the desired property that have larger uncertainty for
points outside the training set.

\begin{figure}
\includegraphics[width=\columnwidth]{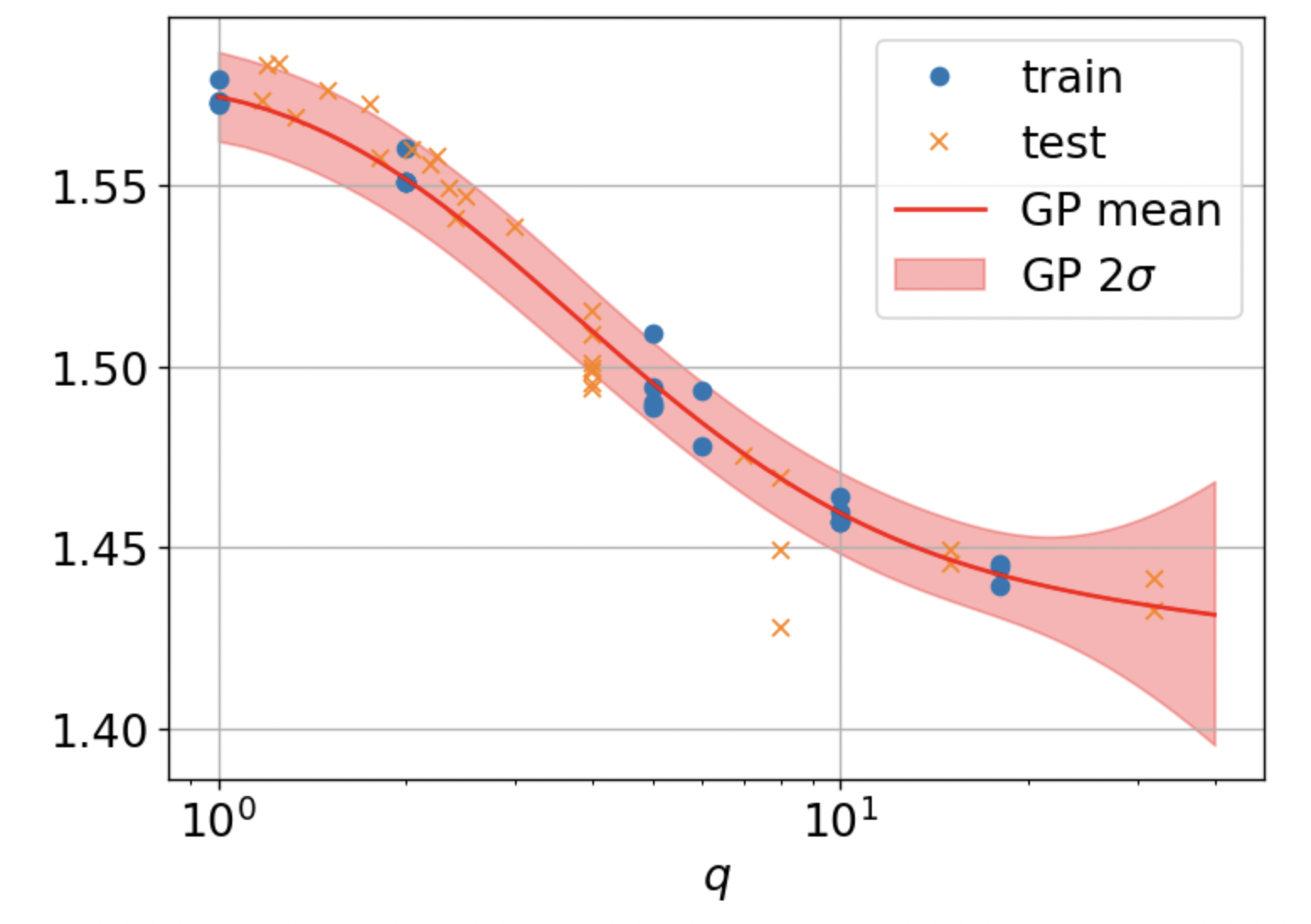}
\caption{Example GPR fit for the amplitude merger model. We show the
data and fit for the collocation point at time $t = 0 M$.
The training and test set are shown as blue and orange points.
The mean GP fit is the red line and the red shaded region
is the $2\sigma$ prediction interval.}
\label{fig:gpr-amp}
\end{figure}

Figure~\ref{fig:gpr-amp} shows the GPR fit
for the peak amplitude (collocation point at $t=0M$).
The blue and orange points are the training and test sets respectively.
The red line is the GPR mean, the red shaded region is
the $2\sigma$ predictive interval.
As we extrapolate the GP model towards mass-ratio 32:1 (the largest mass-ratio simulation
in the test set) we find that the mean prediction agrees well however, the uncertainty
also grows.

\section{Final Model}
\label{sec:fm}

As a reminder, we model the amplitude and angular frequency of the $h_{22}$
complex multipole using a set of linear ans{\"a}tze. The free parameters (collocation
point values) are modelled independently as a probability distribution that
depends on the mass-ratio using GPR.  We denote the GP fit to collocation point
value $\theta$ belonging to either the amplitude (A) or angular frequency ($\omega$) as
$g_{\theta^{[A/\omega]}}(q)$. Using this, we define the final model for the
complex multipole as

\begin{eqnarray}
\hspace{-0.5cm} p(\hat{h}_{22}|t, q) &\equiv& p(\hat{h}_{22}|t, g_{\theta}(q)) \\
   &=& \hat{A}_{IMR}(t, g_{\theta^A}(q)) \exp\left[{-i \hat{\phi}_{IMR}(t, g_{\theta^\omega}(q))}\right] \, .
\end{eqnarray}

The model for the GW phase $\hat{\phi}_{IMR}$ is obtained by generating $\hat
\omega_{IMR}$ first and then numerically integrating it however, an analyitc
expression could be derived. $\hat{A}_{IMR}$ and $\hat \omega_{IMR}$ are given
by Eq. \ref{equ:amp_imr} and Eq. \ref{equ:omega_imr} respectively.

The GPR method provides an analytic expression for the
mean of the GP which we will denote as $\bar{\theta}(q)$.
The mean prediction (which could also be called the best-fit prediction)
is obtained when we use the mean of each GP fit as the
estimate for the collocation values. We define this as

\begin{equation}
\bar{h}_{22}(t, q) \equiv p(\hat{h}_{22}|t, g_{\bar{\theta}}(q)) \, .
\end{equation}

The model can produce independent waveform realisations, denoted by $\tilde{h}$,
by drawing a random sample from the posterior probability for each $\theta$.  We
define this as

\begin{equation}
\tilde{h} \sim p(\hat{h}_{22}|t,g_{\theta}(q)) \, .
\end{equation}

The ability to draw waveform samples can be utilised in Bayesian parameter estimation
of GW events in order to marginalise the posterior over waveform systematic and statistical
uncertainty.

\section{Model validation}
\label{sec:val}

To assess the accuracy between two real-valued time-domain
waveforms $h_1$ and $h_2$ we use the noise-weighted inner product

\begin{equation}
\langle h_1 , h_2 \rangle = 4 \, \text{Re} \int\limits_{f_{\text{min}}}^{f_{\text{max}}} \frac{\tilde{h}_1(f) \tilde{h}_2^*(f)}{S_n(f)} df \, .
\end{equation}

Where $S_n(f)$ is the noise power spectral density of the detector.
The match is defined as the inner product between normalised
waveforms $(\hat{h} = h / \sqrt{\langle h , h \rangle})$
maximised over a relative time and phase shift between $h_1$ and $h_2$.
Our main accuracy metric is the mismatch $\mathcal{M}$ defined as,

\begin{equation}
\mathcal{M}(h_1 , h_2) = 1 - \max\limits_{t_0, \phi_0} \langle \hat{h}_1 , \hat{h}_2 \rangle \, .
\end{equation}

Because the NR data are relatively short (when the total mass is scaled to $100
M_\odot$ the start frequency ranges from $20 - 30$ Hz) we choose to compute the
white-noise mismatch. We do not wish to introduce uncertainty into our results
due to the inability of the NR data to fill the detectors sensitivity band at a
given total mass~\cite{Ohme:2011zm} or introduce an ambiguity into which part
of the waveform is responsible for the error.  In what follows we generate
waveforms with a sample rate of $4096$ Hz scaled to a total mass of $100
M_\odot$.

\begin{figure*}[tbh]
\includegraphics[width=\textwidth]{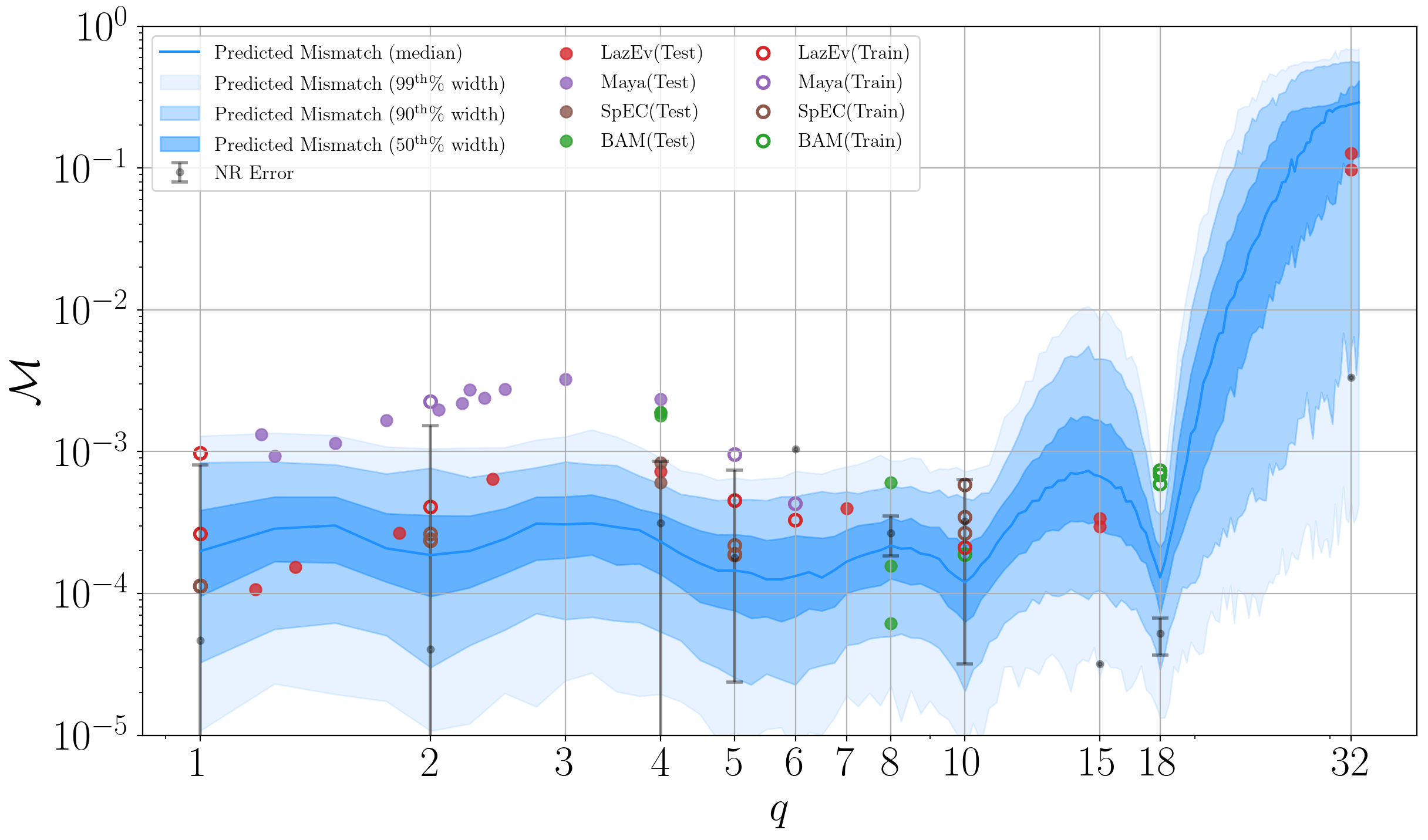}
\caption{The mismatch between various objects as a function of the mass-ratio.
The black error bars are an estimate of the NR uncertainty obtained by
computing the mismatch between NR waveforms at the same mass-ratio.
The mismatch between PPM mean prediction and NR simulations in
the training set (open circles) and the test set (filled circles),
which are coloured based on the NR code,
represent the typical accuracy metric of waveform models.
The solid blue curve is the median of the predicted mismatch distribution
(Eq.~\ref{equ:pm}) and the shaded blue regions show the
$50^{th}$, $90^{th}$ and $99^{th}$ percentile widths.}
\label{fig:mismatch}
\end{figure*}

Figure~\ref{fig:mismatch} shows the results of the mismatch calculation.
We compute the mismatch between the mean PPM waveform and every NR solution in
the train (circle) and test (filled circle) set. The points are coloured
with respect to the NR code used to generate them.

The median mismatch across the test set is $0.13\%$ (a match of $99.87\%$).
The worst mismatch over the test set is $11\%$ (a match of $89\%$) which occurs
for the 32:1 simulation. As this simulation is far away from any training data
and we haven't specifically tuned how the model should extrapolate it is not
surprising that the error is large. The next worst mismatch is $0.32\%$ (a
match of $99.68\%$) which occurs for mass-ratio 3:1. A baseline accuracy
threshold of $1\%$ mismatch error is typically used for which the PPM model
passes for mass-ratios less than 18:1.

To illustrate the uncertainty in the NR waveforms we compute the mismatch
between a reference NR waveform and all other NR waveforms at the same
mass-ratio, these are shown as black circles.  This estimate for the NR error
tends to give error estimates with larger variances when there are more than
one NR code available to compare. For example, simulations at mass-ratios 8:1
and 18:1 are only available with the BAM NR code.  This estimate should be
treated with caution as it assumes all the NR waveforms used in the comparison
are of comparable numerical accuracy which is not true. For instance, we have
included NR waveforms from the same code but performed at different numerical
resolutions.  Future NR simulations available at multiple resolutions that
permit a convergence test would alleviate this issue.

Our PPM can be used to empirically estimate its own uncertainty.  We call this
the predicted mismatch distribution $p_{\mathcal{M}}$ and it is computed as the
distribution of the mismatch between the mean PPM waveform ($\bar{h}^{PPM}$)
and 1000 samples ($\tilde{h}^{PPM}$) from the PPM model.

\begin{equation}
    p_{\mathcal{M}} = \mathcal{M}(\bar{h}^{PPM}, \tilde{h}^{PPM})\, .
\label{equ:pm}
\end{equation}

In Figure~\ref{fig:mismatch} we show the median of $p_{\mathcal{M}}$ as a solid
blue line as well as the $50^{th}$, $90^{th}$ and $99^{th}$ percentile widths
as shaded blue regions.  The results show that the model predicts its error to
be relatively constant between mass-ratio 1:1 and 10:1 at the level of
$0.019_{-0.015}^{+0.039} \%$.  The behaviour between 10:1 and 18:1 suggests
that the uncertainty estimate for the 18:1 is too small resulting in a GP model
that is too heavily constrained in the vicinity of the 18:1 data and can cause
the observed high variance predictions.  Using our prediction for the expected
mismatch we state that we expect the PPM model to likely (at the $99^{th}\%$)
still be accurate at approximately the $1\%$ level when extrapolated to
mass-ratio 20:1.  In the next section we will quantify the accuracy of this
estimate.

There are a number of simulations in the test set, at low mass-ratio between
1:1 and 4:1, that have unusually high mismatches when compared with the PPM
mean model as well as lie outside the predicted mismatch distribution.
Simulations at nearby mass-ratios are available from different NR codes and the
difference seen suggests that the Maya simulations in this region and the BAM
4:1 simulations have numerical errors larger than the other NR codes in our
dataset.  This highlights the potential benefits from constructing training
sets from multiple NR codes to avoid building models that inherit potential
systematic biases from particular NR simulations.

\begin{figure*}[tbh]
\includegraphics[width=\textwidth]{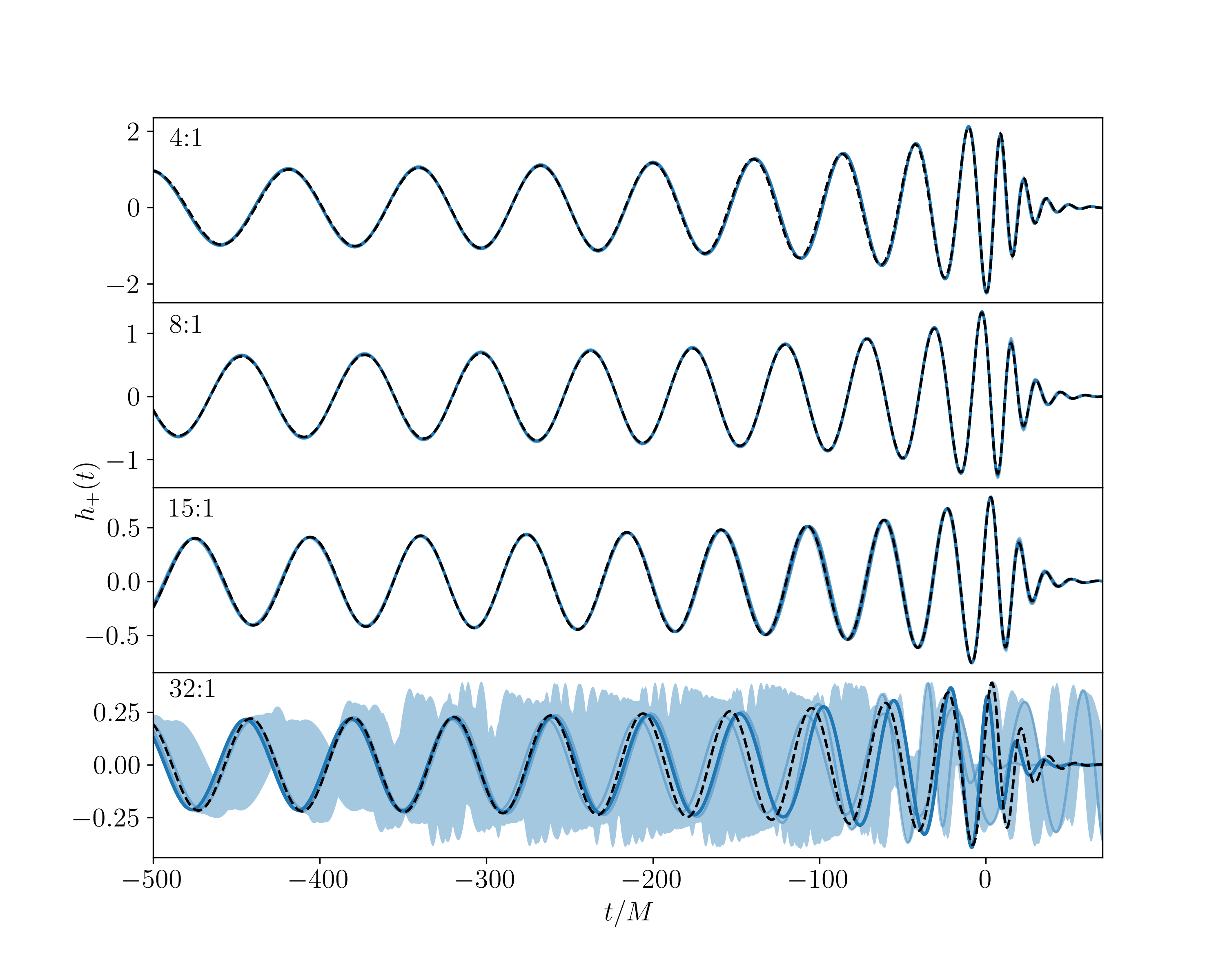}
\caption{In each panel we compare our model with representatative NR simulations at
mass-ratios 4:1. 8:1, 15:1 and 32:1 from the test set.
The NR is shown as a solid black line. We generate 1000 samples and the mean prediction
from our PPM model and optimally align them with the NR waveform over a time and phase shift.
We show the minimum and maximum range of values from the 1000 samples
as the shaded region and explicitly plot the mean as well as 3 samples.}
\label{fig:waveform_plot}
\end{figure*}

Next we compare the PPM model predictions with NR waveforms in the test set to
illustrate how the diversity in waveform predictions under different levels of
uncertainty. Figure~\ref{fig:waveform_plot} shows the $h_+$ waveform from NR
(black, dashed) and predictions from PPM (blue). Both the mean and 3 samples
from PPM are shown as well as the minimum and maximum values from 1000 samples
are shown as the shaded region. From top to bottom we show mass-ratios 4:1,
8:1, 15:1 and 32:1, the mismatch between the mean PPM prediction and NR is
$0.18\%$, $0.016\%$, $0.032\%$ and $11\%$ respectively. The self-mismatch error
for 4:1 and 8:1 are both $0.05\%$ (worst mismatch at $90^{th}$ percentile),
this level of variance in mismatch corresponds to practically indistinguishable
predictions between PPM samples on this scale. As the mass-ratio increases
visible differences begin to be noticeable at mass-ratio 15:1 where the
self-mismatch error gets to the $0.5\%$ (worst mismatch at $90^{th}$
percentile) level. For mass-ratio 32:1 the samples from the PPM model are very
diverse which gives rise to a large mean predicted mismatch ($28\%$) and a wide
distribution with mismatches reaching up to $57\%$ (worst mismatch at $90^{th}$
percentile) for the predicted mismatch distribution.

\section{Uncertainty Calibration}
\label{sec:uncert}

In the previous section we have shown how probabilistic models can be used to
estimate their uncertainty with the use of the predicted mismatch distribution.
However, having the ability to generate waveform samples does not guarantee
that the resulting distribution of waveforms will accurately represent the true
uncertainty of the model.  In this section we quantify the accuracy of our
uncertainty prediction.

We compare the estimated uncertainty with the true uncertainty of the model at
the train and test locations to quantity the accuracy of the uncertainty estimate.  For
the true uncertainty we use the mismatch between NR waveforms and the PPM mean
waveform. We summarise the results with a metric we call the calibration score
$\mathcal{C}$ defined as the ratio between the true uncertainty and the
estimated uncertainty (both measured in terms of the mismatch)

\begin{equation}
\mathcal{C} = \frac{\text{True Uncertainty}}{\text{Estimated Uncertainty}}
\end{equation}

For the estimated uncertainty we use the predicted mismatch distribution
$p_{\mathcal{M}}(q)$ (Eq.~\ref{equ:pm}) from the previous section to obtain a distribution
for the calibration score

\begin{equation}
    p_{\mathcal{C}}(q) = \frac{\mathcal{M}(\bar{h}^{PPM}, h^{NR})}{p_{\mathcal{M}}(q)} \, .
\label{equ:calib}
\end{equation}

A perfectly calibrated model will have $\mathcal{C} = 1$.
A model that is underestimating the uncertainty and is therefore overconfident
will have $\mathcal{C} > 1$, here samples from the PPM will be
closer to the mean prediction than they should be.
A model that is overestimating the uncertainty and is therefore underconfident
will have $\mathcal{C} < 1$, here samples from the PPM will be
further from the mean prediction than they should be.

A previous study~\cite{PhysRevD.96.123011} that also used GPR in waveform
modelling proposed to use the maximum mismatch between the mean waveform and
waveform samples to estimate the true uncertainty i.e. $\max p_{\mathcal{M}}$.
This typically results in estimates of the calibration score that are biased
towards being underconfident.

\begin{figure*}[tbh]
\includegraphics[width=\textwidth]{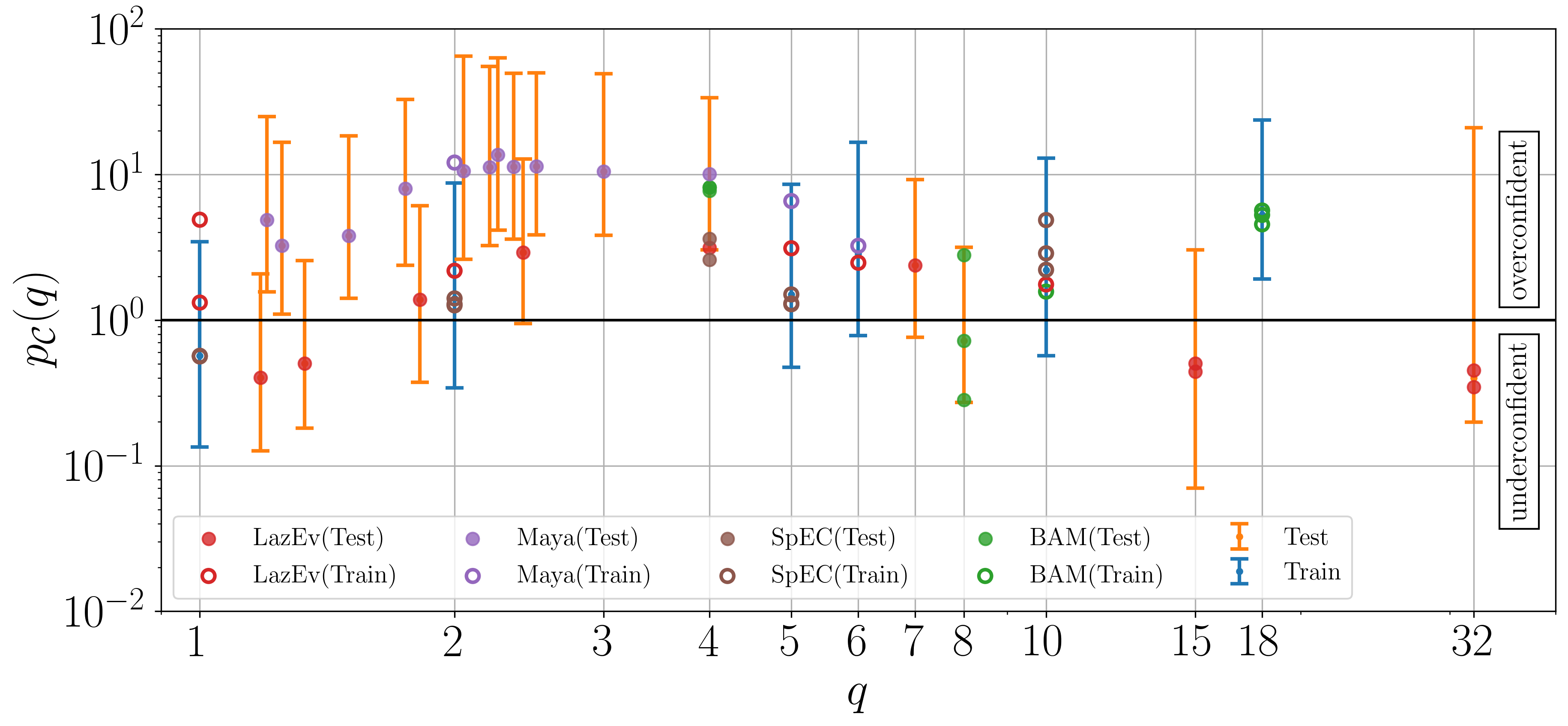}
\caption{Calibration score distribution as defined by Equation~\ref{equ:calib}.
We show the median value and the $90\%$ width of the predicted mismatch
distribution for the train (blue) and test (orange) sets.
We also show the individual results for each NR simulation
compared with the median value of the predicted mismatch.}
\label{fig:calib_plot}
\end{figure*}

The calibration score as a function of the mass-ratio is shown in
Figure~\ref{fig:calib_plot}.  For each mass-ratio where we have more than one
NR simulation we aggregate the results and show the median as well as the
$90\%$ width of the predicted mismatch distribution.  These results are shown
as blue and orange points for the train and test sets respectively.  We also
show $\mathcal{C}$ for each individual NR simulation computed using the median
value of the predicted mismatch distribution.  These results are shown as
circles and filled-circles for the train and test sets respectively.

For the train set we find that the calibration score is consistent with 1 at
the $90 \%$ level for all mass-ratios except the 18:1 simulations.  Here the
model is overconfident in its predictions by a factor of 5 on average.
For the test set we find our model produces
calibrated uncertainties for all cases at the $90 \%$ level except the
mass-ratio 4:1 and all Maya simulations between mass-ratio 1:1 and 4:1. For
these cases the model is consistently overconfident in its uncertainty
estimate.

Our calculation for the calibration score is potentially corrupted due to data
quality issues with the NR data.  In this work we have attempted to control for
this by including as many NR simulations from different codes as possible
however, there simply isn't enough data.  For example, waveforms in the test
set for $q > 1$ and $q < 4$ are mainly from the Maya NR code which could
potentially be a cause of systematic bias in our estimates.  Also we are
treating NR solutions with different numerical resolutions as being equally
accurate.

We hypothesise that the main source of error that is reducing the ability of
our model to accurately predict the true uncertainty is due to data quality
issues which violates our assumption that the NR data are of sufficient and
comparable accuracy\footnote{Recall that we have intentionally
included NR simulations from older catalogues and as such 
they are not necessarily representatative of the accuracy of
current NR codes.}.
The evidence for this is can be seen in
Figure~\ref{fig:mismatch} where the Maya $q \in (1, 4]$ and the BAM $q = 4$
simulations have higher mismatch errors then the other NR simulations when
compared with the PPM model.
Typically these simulations would not be included
in the data set due to data quality concerns however, without a full convergence
series for an NR simulation it is difficult to quantify the errors in a
simulation. However, if the errors in a simulation are quantified
appropriately then these simulations may still add valuable information
but their influence will be downweighted.

\section{Conclusions}
\label{sec:con}

In this paper we address the increasingly important issue of uncertainty
quantification in waveform modelling.  We have presented a new methodology to
build probabilistic phenomenological models (PPMs).  The key aspects of our
work are: (i) employing linear ans{\"a}tze so we can use the collocation
fitting method and gain interpretability, (ii) using a probabilistic fitting
method (such as Gaussian process regression) for the parameter space fits and
(iii) using estimates for the NR uncertainty to inform those fits.  PPMs extend
current phenomenological methods with the ability to not only generate the
best-fit point estimate but also explicit waveform samples that can be used to
marginalise over waveform model errors in GW Bayesian parameter estimation.

The model presented here is a proof of concept. It only covers a small portion
of the waveform ($\sim 800 M$ in duration) and does not model spinning
binaries. It should be relatively straightforward to adapt current methodology
used to build \emph{deterministic} phenomenological models, that model
precessing binaries with higher order multipoles, and turn them into
\emph{probabilistic} phenomenological models.  NR solutions of these more
complete descriptions of binary coalescence typically have larger numerical
errors and therefore stand to benefit the most from a probabilistic treatment.

Some interesting technical challenges have the potential to appear when
increasing the size of the dataset and/or including more physics.  For example,
the noise model assumption may need to be revised if the data shows sign of
heteroskedasticity. Another assumption is the independence of the
phenomenological coefficients. Future models will likely continue using
non-linear ans{\"a}tze that are more physically motivated.
If the coefficients of these models have significant correlation
then our method of sampling the coefficients independently could
result in unphysical waveforms. In such a case then modelling algorithms that
can jointly model the coefficients will have to be explored.

The most important issue that needs to be resolved is to improve the error
estimates in NR simulations as this is a crucial ingredient in modelling.  We
have experiemented with using the difference between NR solutions from
different NR codes to estimate the NR error however, this is not reliable.
Where possible we recommend NR groups publish detailed uncertainty estimates
that are functions of time along with their waveforms.

\begin{acknowledgments}
We thank Cardiff University as well as Gregory Ashton, Deborah Ferguson,
Xisco Jim\'enez-Forteza, Shrobana Ghosh, Mark Hannam, Frank Ohme, Jonathan
Thompson and Prim for useful discussions.
\end{acknowledgments}

\bibliography{paper}

\end{document}